# Does Building a Relative Sunspot Number Make Sense? A Qualified 'Yes'

Leif Svalgaard[1] (leif@leif.org)

[1] Stanford University, Cypress Hall C13, W.W. Hansen Experimental Physics Laboratory, Stanford University, Stanford, CA 94305, USA

**Abstract:**

Recent research has demonstrated that the number of sunspots per group ('active region') has been decreasing over the last two or three solar cycles and that the classical Relative Sunspot Number (SSN) no longer is a good representation of solar magnetic activity such as revealed by *e.g.* the F10.7 cm microwave flux. The SSN is derived under the assumption that the number of spots per group is constant (in fact, nominally equal to 10). When this is no longer the case (the ratio is approaching 5, only half of its nominal value) the question arises how to construct a sunspot number series that takes that into account. We propose to harmonize the SSN with the sunspot Group Count that has been shown to follow F10.7 very well, but also to include the day-to-day variations of the spot count in order to preserve both long-term and short-term variability.

## 1. Introduction

Johann Rudolf Wolf's observation, almost by happenstance, on December 4th, 1847 of a large sunspot (wolf, 1856) excited an enduring (46+ year) interest in the sunspot phenomenon, its observation, and quantitative description. The discovery by Heinrich Schwabe of the sunspot cycle and the discovery by Wolf himself (Wolf, 1852) and, independently, by Gautier (Gautier, 1852) that the amplitude of the diurnal variation of the geomagnetic Declination (angle between compass needle and true North) seemed to vary in step with the newly discovered sunspot cycle gave further impetus to the observations and to that study of the cycle, which would last for the rest of Wolf's life. Today, the sunspot record initiated by Wolf is often the primary input to reconstructions of various aspects of solar activity used in both solar and climate research (*e.g.* Krivova *et al.*, 2010).

Wolf started his regular observing program at Bern, Switzerland in 1849 using a 4-foot refractor at magnification 64 manufactured by Fraunhofer. Wolf recorded for each day, on which observations were made, two numbers: the first giving the number of sunspot groups and the second the total number of single spots contained in all groups. All observations were recorded in this same basic format and published each year until 1945 when the publication of those 'raw' data unfortunately was discontinued by Waldmeier, when he was taking over the directorship of the Zürich Observatory. In order to compile monthly and yearly values of the observations, Wolf formed his famous daily Relative Sunspot Number, $R$, as 10 times the number of groups, $g$, plus the total number of spots, $f$, so that $R = 10\,g + f$. The formation of a new group is clearly a much more important event than the formation of yet one more spot within an existing group, so giving the number of groups a high weight captures that importance. The specific weight '10' emerged from a combination of experience (that a group on average contains about 10 spots) and convenience. The combination with equal contributions of the weighted group count and the spot count makes for a very 'robust' index that has abundantly proven its usefulness over time to this day.

Later Wolf (Wolf, 1861) introduced a 'scale' factor $R = k\,(10g + f)$ to enable observations by observers using different instruments, different selection criteria, and having different Snellen ratios (acuity) to be brought on to the same scale, namely his own (so $k=1$ for Wolf). To be compatible with Heinrich Schwabe's observations, Wolf decided not to count the smallest spots at the limit of visibility. His successor Alfred Wolfer (Wolfer, 1894) argued that this criterion was too vague and proceeded (after Wolf's death in 1893) to count all visible spots, necessitating the use of a smaller *k*-factor (0.6) to bring his counts down onto Wolf's scale. This has given rise to endless confusion (*e.g.* to the 'deduction' that an average group contains 2 spots: Group Sunspot Number = 12 times Zürich Sunspot Number = 10 times Groups + Spots, giving Spots = 2 for Groups = 1). To be useful and user friendly, the Sunspot Number should be freed from all such 'artificial' encumbrances, including the increased 'weighting' of large sunspots, introduced by Waldmeier in ~1947 (Svalgaard and Cagnotti, 2015) and the increased group count due to employment of the Waldmeier Group Classification scheme from ~1940 (section 4.1 of Svalgaard and Schatten, 2015).

## 2. A 'Correct' Relative Sunspot Number?

Is there such a thing as the 'correct sunspot number'? I believe there is and I shall in this short note explain why I think so. I shall first define what I mean by 'solar activity', namely that which index is above a certain base level of the sun's magnetic field, corresponding to when no active regions and no sunspots are visible on the disk (perhaps over a suitably long time interval). This definition is certainly suitable for many recent decades. Built into Wolf's formula (that makes groups and spots contribute equally) is the assumption that on average the number of spots per group is a constant 10 (for Wolf's standard 80mm aperture telescope). We showed in Svalgaard (2015) that it is possible to reconstruct the EUV flux from the diurnal variation of the geomagnetic field, and that that reconstruction follows the variation of the F10.7 microwave flux very closely. If we define solar activity pragmatically as that which has influence on geospace, the 'sunspot number' should be an index that reflects that influence as nearly as possible. The sunspot group number, GN (Svalgaard and Schatten, 2015), satisfies that requirement and the long-term variation of solar activity appears well represented by GN. But the number of spots, SN, carries additional information on the time scale of days, not well captured by the more slowly varying number of groups, so it will be useful to construct an index incorporating both GN and SN, as the original Relative Sunspot Number did. Such a series, the Wolf Number WN, would also be a natural focal point for correlative studies of the responses of the terrestrial and planetary systems to solar activity, while the 'raw' GN and SN are better suited for studies of the Sun itself.

## 3. The Proposed Wolf Number Series

We put the scale of the Wolf Number (WN) to 20 times the Group Number, derived by dividing the (Hoyt and Schatten, 1998) Group Sunspot Number (GSN) scale factor of 12 by 0.6. As the first step, we remove the effect of weighting the sunspot count since 1947 through the present, using the functional form for the weight factor to apply to the relative sunspot number determined in section 5.2 of Clette *et al.* (2014). For each year, we compute the ratio, $f$ (year) = 20·<GN>/<$SSN_{corr}$> between the yearly averages of 20 times the GN and the corrected for weighting SIDC/SILSO/Zürich Sunspot Number $SSN_{corr}$ (any other SSN that you might prefer would do as well). In the next step, each daily value (that is not missing) of $SSN_{corr}$ is now multiplied by the appropriate $f$-value, WN = $f \cdot SSN_{corr}$. This ensures that the yearly average of WN will match the yearly average of GN (there is a subtlety for missing data) to preserve the long-term variation, and at the same time preserves the short-term variation of the SSN. From the new daily WN-series we can then as usual compute monthly and yearly means, and consider computing 27-day (Bartels) rotation averages as well. Figure 1 shows the monthly and monthly smoothed WN-series since 1818.

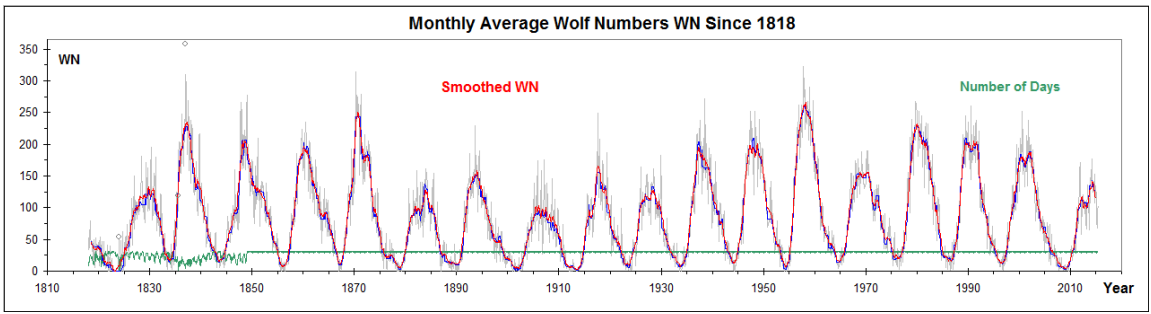

Figure 1: Monthly averages (gray curve) of the Wolf Number series (WN). Outliers based on less than 5 days per months are shown separately by small circles. The monthly smoothed values (using a simple boxcar 12-month smooth) are shown by the red curve. The number of days for each month is shown at the bottom by the green curve. Note the annual variation before 1849 (less data in northern winter).

Text files containing the daily, monthly, Bartels rotation, and yearly values can be found on my website at http://www.leif.org/research/. The formats of these files are as follows:

Yearly file
```
1818  47.42  213
1819  34.49  249
1820  26.37  224
Year   WN    Days
```

Monthly file
```
181801  1818.047  52.00   8
181802  1818.131  33.50  14
181803  1818.214  37.86  14
YYYYMM    Time     WN   Days
```
'-1' means no data

Daily file
```
18180116  1818.045   ?    ?
18180117  1818.048  46   69
18180118  1818.051  59   88
YYYYMMDD   Time    SSN   WN
```
'?' means no data

In the Yearly and Monthly Files, we give the number of days with data. The Daily File gives both the original Sunspot Number (SSN) and the proposed Wolf Number (WN). There is also a Bartels 27-day rotation averages file where the YYYYMMbb is replaced by BBBBBb, which is the Bartels Rotation number; the 'Time' values refer to the middle day of the interval in question.